\documentclass[modern]{aastex7}
\usepackage{CJKutf8}

\newcommand{\K}{\ensuremath{\mathrm{K}}}

\newcommand{\cms}{\ensuremath{\mathrm{cm}\,\mathrm{s}^{-1}}}

\newcommand{\ergs}{\ensuremath{\mathrm{erg}\,\mathrm{s}^{-1}}}

\newcommand{\Ms}{{\ensuremath{{M}_{\odot} }}}

\newcommand{\Zsun}{\ensuremath{\mathrm{Z}_{\odot}}}
\newcommand{\rev}[1]{#1}

\newcommand{\cutt}[1]{\textcolor{blue}{}}

\newcommand{\Ni}{{\ensuremath{^{56}\mathrm{Ni}}}}

\newcommand{\jFe}{{\ensuremath{^{54}\mathrm{Fe}}}}
\newcommand{\iFe}{{\ensuremath{^{52}\mathrm{Fe}}}}

\newcommand{\Hea}{{\ensuremath{^{3} \mathrm{He}}}}
\newcommand{\Heb}{{\ensuremath{^{4} \mathrm{He}}}}
\newcommand{\Hy}{{\ensuremath{^{1} \mathrm{H}}} }
\newcommand{\Ox}{{\ensuremath{^{16}\mathrm{O}}}}
\newcommand{\Ti}{{\ensuremath{^{44}\mathrm{Ti}}}}
\newcommand{\Si}{{\ensuremath{^{28}\mathrm{Si}}}}
\newcommand{\Mg}{{\ensuremath{^{24}\mathrm{Mg}}}}
\newcommand{\Cx}{{\ensuremath{^{12}\mathrm{C}}}}

\newcommand{\Cr}{{\ensuremath{^{48}\mathrm{Cr}}}}
\newcommand{\Ca}{{\ensuremath{^{40}\mathrm{Ca}}}}
\newcommand{\Ar}{{\ensuremath{^{36}\mathrm{Ar}}}}
\newcommand{\Sx}{{\ensuremath{^{32}\mathrm{S}}}}
\newcommand{\Nx}{{\ensuremath{^{14}\mathrm{N}}}}
\newcommand{\Ne}{{\ensuremath{^{20}\mathrm{Ne}}}}

\newcommand{\Figure}[1]{\textbf{Figure~\ref{fig:#1}}}

\newcommand{\Sectff}[1]{{\ref{sec:#1}}}
\newcommand{\Sect}[1]{{\S~\Sectff{#1}}}

\newcommand{\eq}[2]{\begin{equation} \label{eq:#1} #2 \end{equation}}

\newcommand{\CASTRO}{{\texttt{CASTRO}}}
\newcommand{\FLASH}{{\texttt{FLASH}}}

\newcommand{\MESA}{{\texttt{MESA}}}

\newcommand{\mn}[1]{\textbf{\textcolor{black}{R#1}}}

\DeclareRobustCommand{\dl}{\bgroup\markoverwith{\textcolor{red}{\rule[.5ex]{2pt}{0.4pt}}}\ULon}

\begin{document}
\begin{CJK*}{UTF8}{bkai}
\title{\rev{Multi-wavelength Signatures of Supernova Shock Breakout from Red Supergiants in Two Dimensions}}

\author[0009-0002-3816-4732]{Wun-Yi Chen（陳文翊）}
\affiliation{National Taiwan University, Graduate Institute of Astrophysics, Taipei, Taiwan, R.O.C.}
\affiliation{Academia Sinica, Institute of Astronomy and Astrophysics, Taipei 106319, Taiwan, R.O.C.}
\affiliation{Heidelberger Institut für Theoretische Studien, Schloss-Wolfsbrunnenweg 35, 69118 Heidelberg, Germany}
\email{wychen@asiaa.sinica.edu.tw}

\author[0000-0002-4848-5508]{Ke-Jung Chen}
\affiliation{Academia Sinica, Institute of Astronomy and Astrophysics, Taipei 106319, Taiwan, R.O.C.}
\affiliation{Heidelberger Institut für Theoretische Studien, Schloss-Wolfsbrunnenweg 35, 69118 Heidelberg, Germany}
\email{chenken1229@gmail.com}
\author[0000-0003-2611-7269]{Keiichi Maeda}
\affiliation{Department of Astronomy, Kyoto University, Kitashirakawa-Oiwake-cho, Sakyo-ku, Kyoto 606-8502, Japan}
\email{keiichi.maeda@kusastro.kyoto-u.ac.jp}
\author[0000-0002-0603-918X]{Masaomi Ono}
\affiliation{Academia Sinica, Institute of Astronomy and Astrophysics, Taipei 106319, Taiwan, R.O.C.}
\affiliation{Astrophysical Big Bang Laboratory (ABBL), RIKEN Pioneering Research Institute (PRI), \\
2-1 Hirosawa, Wako, Saitama 351-0198, Japan}
\email{masaomi@asiaa.sinica.edu.tw}
\author[0000-0003-1295-8235]{Po-Sheng Ou（歐柏昇）}
\affiliation{Academia Sinica, Institute of Astronomy and Astrophysics, Taipei 106319, Taiwan, R.O.C.}
\affiliation{Department of Physics, National Taiwan University, No.1, Sec. 4, Roosevelt Rd., Taipei 10617, Taiwan, R.O.C.}
\email{psou@asiaa.sinica.edu.tw}
\author[0000-0002-4460-0097]{Friedrich K.\ R{\"o}pke}
\affiliation{Zentrum für Astronomie der Universität Heidelberg, Astronomisches Rechenzentrum, M{\"o}nchhofstr.\ 12--14, 69120 Heidelberg, Germany}
\affiliation{Heidelberger Institut für Theoretische Studien, Schloss-Wolfsbrunnenweg 35, 69118 Heidelberg, Germany}
\affiliation{Zentrum für Astronomie der Universität Heidelberg, Institut für Theoretische Astrophysik, Philosophenweg 12, 69120 Heidelberg, Germany}
\email{friedrich.roepke@h-its.org}
\begin{abstract}

We present new two-dimensional radiation hydrodynamic simulations of supernova shock breakout from red supergiants using the {\CASTRO} code. Our progenitors are 20 and 25 \Ms\ solar-metallicity stars evolved from the zero-age main sequence with {\MESA} and exploded in one dimension using {\FLASH}. We consider a range of circumstellar media (CSM) produced by stellar winds to investigate how pre-explosion mass-loss affects shock breakout. The multigroup flux-limited diffusion scheme in \CASTRO\ captures the interaction between the explosion shock, its radiation precursor, and the surrounding CSM. We find that strong radiation precursors, generated by radiation leakage behind the shock, can drive fluid instabilities and move the effective photosphere outward before the shock reaches the stellar surface. The resulting breakout emissions reach peak luminosities of ${\sim}10^{44}$~\ergs\ with full width at half maximum durations of 1--3~hr\rev{, fainter and longer than previous 1D models.} The light-curve colors gradually evolve from blue to red after the peak. The 25 \Ms\ model with explosion energy $E \sim 1.69\times10^{51}$~erg produces ${\sim}$10--30\% higher maximum luminosity than the 20 \Ms\ model with $E \sim 1.09\times10^{51}$~erg. The dense CSM further extends the breakout rise time by increasing the photon diffusion. These results provide new constraints on red supergiant atmospheres and mass-loss histories prior to core collapse.
\rev{Although our selected wavelength ranges fall outside the observational bands of missions such as Einstein Probe and ULTRASAT, our study still provides valuable insights into the color evolution of shock breakout and can help guide the interpretation of future observations.}

\end{abstract}
\section{Introduction} \label{sec:intro}
The first electromagnetic signal from a supernova (SN) emerges when the shock approaches the stellar surface with its optical depth reduced to $\tau \sim \frac{c}{v_{\rm s}}$, where $c$ is the speed of light and $v_{\rm s}$ is the velocity of the shock. The shock breakout of SNe encodes key information about their progenitor stars, the explosion mechanisms, and surrounding environments, which are available only during the earliest phase of explosion \citep[see, e.g.,][]{2008Schawinski, 2011Couch, 2015Gezari, 2017Waxman}. Detections of this transient phenomenon become increasingly promising through direct observations from wide-field X-ray telescopes on timescales of ${\gtrsim}0.1$ hours (hr) and UV follow-up observations lasting up to a day \citep{2022Bayless, 2023Hosseninzadeh, 2024Shrestha}, as well as indirect methods using light echoes and gravitational lensing \citep{2008Dwek, Suwa2017lensing}. Depending on the progenitor radius and circumstellar medium (CSM) density, shock breakout signals from red supergiants (RSGs) can reach ${\sim} 10^{44}$~\ergs\ and last for several hours to days, making them easier to detect than shock breakouts from blue supergiants (BSGs) and Wolf-Rayet stars (WRs), which only last for several tens of minutes \citep{2008Soderberg, suzuki2016,2020RSG_Alex,rsgwind}.
 
To explain the prolonged shock breakout signatures observed in some SNe \citep{2011Chevalier,2015Gezari,2017Yaron}, a dense CSM in the vicinity of the progenitor is often required, as supported by previous one-dimensional (1D) studies \citep{2017Lovergrove,2017Dessart,2022Kozyreva} \rev{and recently also by three-dimensional (3D) simulations of envelope pulsations in RSGs \citep{2025arXivMa} as well as yellow supergiants \citep{2025Goldberg}}. A sufficiently dense CSM provides large optical depths that regulate the emission from CSM breakouts, enabling shock breakout light curves (LCs) to last for several hours. Furthermore, recent 3D simulations of shock breakouts based on 3D RSG envelope structure have shown that clumpy atmospheric structures can increase photon diffusion times, thereby prolonging the breakout LC \citep{2022Goldberg}. 
Together, the clumpy envelope and dense CSM shift the energetic photons emerging from the shock breakout of RSGs to longer wavelengths, producing radiation that spans from the infrared to soft X-rays \citep{Katz_2010,2013Tolstov,2022Margalit}.

Multidimensional and multigroup RHD simulations are required to model the complicated coevolution of gas and radiation during shock breakout in RSGs, where strong mixing occurs during the breakout and shock-CSM interaction phases \citep{Katz_2010, 2024Chen}. For example, our previous work \citep{2024Chen} presented the first two-dimensional (2D) multigroup RHD simulations for SN 1987A, demonstrating the dimensional effect and color evolution of the shock breakout. However, our previous models are not directly applicable to RSGs, whose stellar structure differs from that of the BSG used in our SN 1987A study. \rev{Meanwhile, multidimensional RHD studies of RSG shock breakout either employ grey radiation transport and cannot track color evolution (e.g., \citealt{2022Goldberg}) or estimate color light curves only after breakout in long-term 3D simulations \citep{2025Vartanyan}.} Therefore, we extend multigroup RHD modeling to RSGs using 2D simulations with {\CASTRO} \citep{2010Almgren, Almgren2020} to advance our understanding of shock breakout for RSGs.

We first introduce the progenitor models and numerical methods in \Sect{NM}. Then, we present the simulation results in \Sect{results}, and discuss their astrophysical implications in \Sect{dis}. Finally, we conclude with a summary of our findings in \Sect{conclusions}.

\section{Numerical Method}
\label{sec:NM}
\subsection{\CASTRO}
We perform our 2D multi-wavelength radiation hydrodynamics (RHD) simulations of RSG shock breakout using the \CASTRO\ code.
The {\CASTRO} code \citep{2010Almgren, Almgren2020} is a public code designed for astrophysical simulations, and it solves the equations of compressible RHD using a higher-order Godunov scheme for hyperbolic radiation fluid and frequency-space advection. Photon diffusion and source-sink terms are treated with a first-order backward Euler method. The code employs a co-moving frame multigroup flux-limited diffusion (MGFLD) formulation \citep{2011Zhang, 2013Zhang}, and includes self-gravity through a Poisson solver \citep{2016Katz}. {\CASTRO} supports adaptive mesh refinement (AMR) to resolve the fine structure of fluid instabilities emerging in astrophysical flows.

The RHD module in {\CASTRO} evolves gas and radiation temperatures independently; the gas and radiation temperatures within a given cell can diverge and allow for a non-local thermal equilibrium. We use an approximate flux limiter from \citet{1981Levermore} to close the radiation momentum equation, relating the radiation flux to the local optical depth. Our frequency groups are logarithmically spaced and range from ${\sim} 10^{14}$ Hz to ${\sim} 3 \times 10^{17}$ Hz, corresponding to the photon wavelength that spans from ${\sim} 10^5$~\text{\AA}\ (infrared) to $10$~\text{\AA}\ (X-ray). The exact range is model-dependent and follows the group selection strategy of \citet{2017Lovergrove} based on the evolution of the radiation energy.

Electron scattering dominates the opacity during shock breakout; therefore, our model assumes temperature-independent opacity $\kappa_{\rm e}=0.32\,\rm{(cm^{2}\,g^{-1})}$ with assumptions of electron fraction $Y_{\rm e}=0.85$. We assume that the inelastic scattering fraction $f=10^{-4}$ contributes to absorption $\kappa_{\rm a}=f\times \kappa_{\rm e}$ \citep{2024Chen}, suggested by \citet{2013Zhang,2017Lovergrove}. \rev{We adopt an ideal-gas equation of state with a constant adiabatic index $\gamma=5/3$ and evolve radiation separately using MGFLD. Ionization and recombination effects are not included.}

\subsection{Progenitor Stars}
\label{subsec:progenitor}
\subsubsection{1D Stellar and Supernova Models}
Our 1D RSG progenitors take 20 \Ms\ and 25 \Ms\ solar-metallicity (\Zsun) stars from \cite{Ou_2023}, which contain a large grid of stellar models with {\MESA} \citep{Paxton2011, Paxton2013, Paxton2015, Paxton2018, Paxton2019}. These progenitor masses are broadly consistent with previous studies on the explosion of RSG shocks, including the 15 and 25 \Ms\ models in \cite{2017Lovergrove} and the ${\sim}$19 \Ms\ model in \cite{2022Goldberg}. Our progenitors approach the upper mass limit of typical RSG progenitors found for type II SNe \citep{2022Rodriguez}, although this boundary remains a topic of debate in stellar evolution theory and observational studies \citep{2009Smartt, 2024ChenKaitlyn,2025Beasor}.

\begin{figure}
    \centering
    \includegraphics[width=1\linewidth]{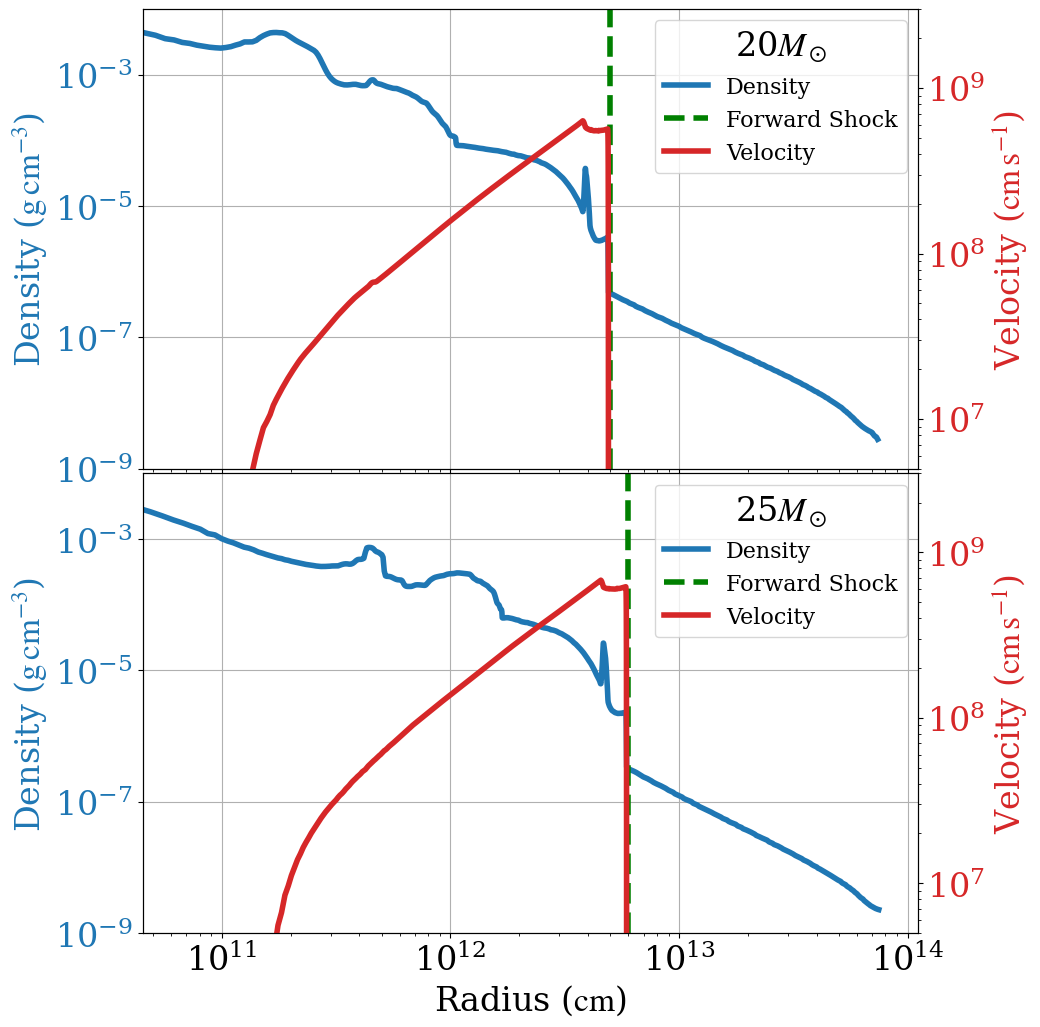}
    \caption{1D density and radial velocity profiles for \mn{20} and \mn{25} when the shock reaches the hydrogen envelope. At this moment, the corresponding shock velocity is $v=6.35\times 10^{8}$~\cms\ for \mn{20} and $v=6.8\times 10^{8}$~\cms\ for \mn{25}.
    The original stellar radii of \mn{20} and \mn{25} are $R_{\rm *}=7.6\times 10^{13}$ cm and $R_{\rm *}=8.3\times 10^{13}$ cm, respectively.}
    \label{fig:1}
\end{figure}

The 20 \Ms\ star evolved to a final mass of ${\sim}$16 \Ms\ with a radius of $7.69 \times 10^{13}$ cm \rev{(1105~$R_{\odot}$)} after ${\sim}$8 Myr and the 25 \Ms\ star evolved to a final mass of 17 \Ms\ with a radius of $8.34 \times 10^{13}$ cm \rev{(1200~$R_{\odot}$)} after 6.4 Myr, before they died. Once the star's iron core starts to collapse, we explode the star with the 1D {\FLASH} code \citep{2000Fryxell}. Similarly to \citet{2020ono,2024Ono}, the explosion is initiated by instantaneously injecting thermal and kinetic energies around the interface between the iron core and the silicon-rich layer of the progenitor star. The bulk iron core is assumed to collapse into a neutron star, treated as a point-like gravity source. As the SN shock propagates, the computational domain (covered by uniform 1024 cells) is gradually expanded to capture the shock front and the ejecta structure. We use the 19-isotope APPROX reaction network to follow nuclear burning \citep{1995WoosleyWeaver,1999Timmes}. This network evolves mass fractions of \Hy, \Hea, \Heb, \Cx, \Nx, \Ox, \Ne, \Mg, \Si, \Sx, \Ar, \Ca, \Ti, \Cr, \iFe, \jFe, \Ni, protons, and neutrons. It includes alpha-chain reactions, a heavy-ion reaction network, hydrogen burning cycles, photodisintegration of heavy nuclei, and energy loss due to thermal neutrinos.
The explosion energies with the synthesized {\Ni} masses for the 20 \Ms\ and 25 \Ms\ stars are $1.09 \times 10^{51}$~erg\ ($4 \times 10^{-2}\,\Ms$ \Ni) and $1.69 \times 10^{51}$~erg\ ($1.58 \times 10^{-1}\,\Ms$ \Ni), respectively.
We follow the shock propagation until it reaches beyond the $\mathrm{H}$/$\mathrm{He}$ shell boundaries as shown in \Figure{1}.

Next, we map the resulting 1D \FLASH\ profiles onto 2D grids of \CASTRO, with the assumption of black-body spectra and local thermal equilibrium. We then connect the progenitor to the CSM and apply velocity perturbations with amplitude ${\sim}$30\% of the local fluid speed within the stellar envelope to mimic convective motions \citep{2013Ken}. This perturbation amplitude is comparable to previous studies that introduced ${\sim} 30\%$ velocity perturbations \citep{1998Nagataki} or ${\sim}$25~\% density perturbations \citep{Mao_2015} to model mixing in SN explosions. Although these envelope perturbations trigger fluid instabilities, their impact on the resulting LCs is minor compared to the breakout dynamics. Furthermore, we set the perturbation in the CSM to zero because its origins are highly uncertain. 

Although our 1D progenitor stars and their explosion profiles are globally spherical, neglecting multidimensional effects during the core-collapse and early explosion phases \citep[e.g.,][]{suzuki2016,2022Nakamura,2024Wang}, the subsequent explosion shock modeled in 2D \CASTRO\ runs can drive strong mixing and generate significant multidimensional features.

The size of our 2D cylindrical box is $10^{15}$ cm on the $r$ axis and $2 \times10^{15}$ cm on the $z$ axis, resolved with fixed grids of $4096 \times 8192$. \rev{We apply reflective boundary conditions along the symmetry at $r=0$ ($z$ axis) and outflow boundary conditions elsewhere. Our 2D simulations adopt cylindrical symmetry, in which the 3D space is mapped onto a 2D domain by eliminating the azimuthal dimension. In this setup, symmetry about the $z$-axis must be preserved, and we therefore impose reflective boundary conditions to properly model the stellar interior. The remaining outer boundaries are set to outflow to allow the ejecta to freely expand following the explosion. We evolve each model for ${\sim}$48--72~hr, following the shock as it propagates through the hydrogen envelope, breaks out of the stellar surface, and produces rising and fading LCs prior to the onset of hydrogen recombination of envelope.}

\subsubsection{Stellar winds from RSGs}
We assume a fiducial RSG wind mass-loss rate $\dot{M}\sim6\times10^{-6}\,\Ms\,\rm{yr^{-1}}$ with the constant wind velocity of $v_{\rm w}\sim3.6\times 10^{6}\,\cms$ \citep{rsgwind} and calculate the corresponding CSM density based on the radial distance $r$ from the star:
\eq{1}{\rho(r)=\frac{\alpha\dot{M}}{4\pi\,r^2\,v_{\rm w}}\rm{,}}
where $\alpha$ is a scaling factor. By adopting $\alpha$ of 1.0 and 5.0 times the fiducial rate, we investigate the impact of CSM on SN shock breakout \citep{2014Smith}. We note that the actual CSM structure around RSGs can be complicated and contain a densely confined shell or clumpy disk, as suggested by observations \citep{2017Yaron, Hiramatsu_2023, 2025vanLoon}. \citet{2017Dessart} used 1D RHD simulations to study shock breakout through a dense CSM, reproducing the narrow spectral lines in SN~2013fs. They suggested that prolonged breakout emission could result from an extended atmosphere or slow wind, although the outcome depends sensitively on the density gradient between the envelope and inner CSM, as well as radiation interaction within. Since 1D models cannot capture the mixing of shock–CSM collisions, multidimensional simulations are essential to resolve the resulting structures and emission features \citep{2024Chen}.

We summarize the model naming: \mn{20}, \mn{25}, \mn{20T}, and \mn{25T} in Table~\ref{tab:name}, where \textbf{R} stands for \textbf{R}SG and the numbers represent the progenitor ZAMS mass. The capital letter \textbf{T} represents a \textbf{T}hick CSM.
\begin{deluxetable*}{lcc}
\tabletypesize{\scriptsize}
\tablewidth{0pt}
\tablecaption{Model Summary \label{tab:name}}
\tablehead{
\colhead{Model} & \colhead{Progenitor star mass } & \colhead{CSM density via $\alpha$}
} 
\startdata 
{ \mn{20}}&  $20\,\Ms$   & 1.0 \\
{ \mn{25}}&  $25\,\Ms$   & 1.0 \\
{ \mn{20T}}& $20\,\Ms$   & 5.0 \\
{ \mn{25T}}& $25\,\Ms$   & 5.0 \\
\enddata
\tablecomments{$\alpha$ is the scaling parameter for the CSM density profile determined by the stellar wind based on Equation~(\ref{eq:1}).}
\end{deluxetable*}
\subsection{Light Curve Calculations}
To retrieve the observational signatures, we calculate the color LCs based on our RHD simulations following the method of \citet{2024Chen}. Specifically, we extract the frequency-dependent radiation flux ($F_\nu$) and compute the effective luminosity as $L_\nu = 4\pi r^2 F_\nu$ for a distant observer located at $r > r(\tau = 2/3)$ based on the flux at $r\approx 2\times 10^{14}$ cm. We calculate angle-dependent LCs at five viewing angles: $15^\circ$, $30^\circ$, $45^\circ$, $60^\circ$, and $75^\circ$, assuming cylindrical symmetry to project the 2D data onto a pseudo-3D photosphere. Light-travel-time corrections (LTTCs) are applied to each emitting surface element and smooth the LCs. In general, LCs vary with viewing angles, and we select LCs with a viewing angle of $45^{\circ}$ for model comparison. The duration of the LC is defined as the full width at half maximum (FWHM) of the luminosity, consistent with observational interpretations \citep{2011Chevalier} and the shock crossing time \citep{2022Goldberg}.
\subsection{Effect of Atmosphere-CSM Density Gradient}
\label{subsec:verify}
\begin{figure}
    \centering
    \includegraphics[width=1\linewidth]{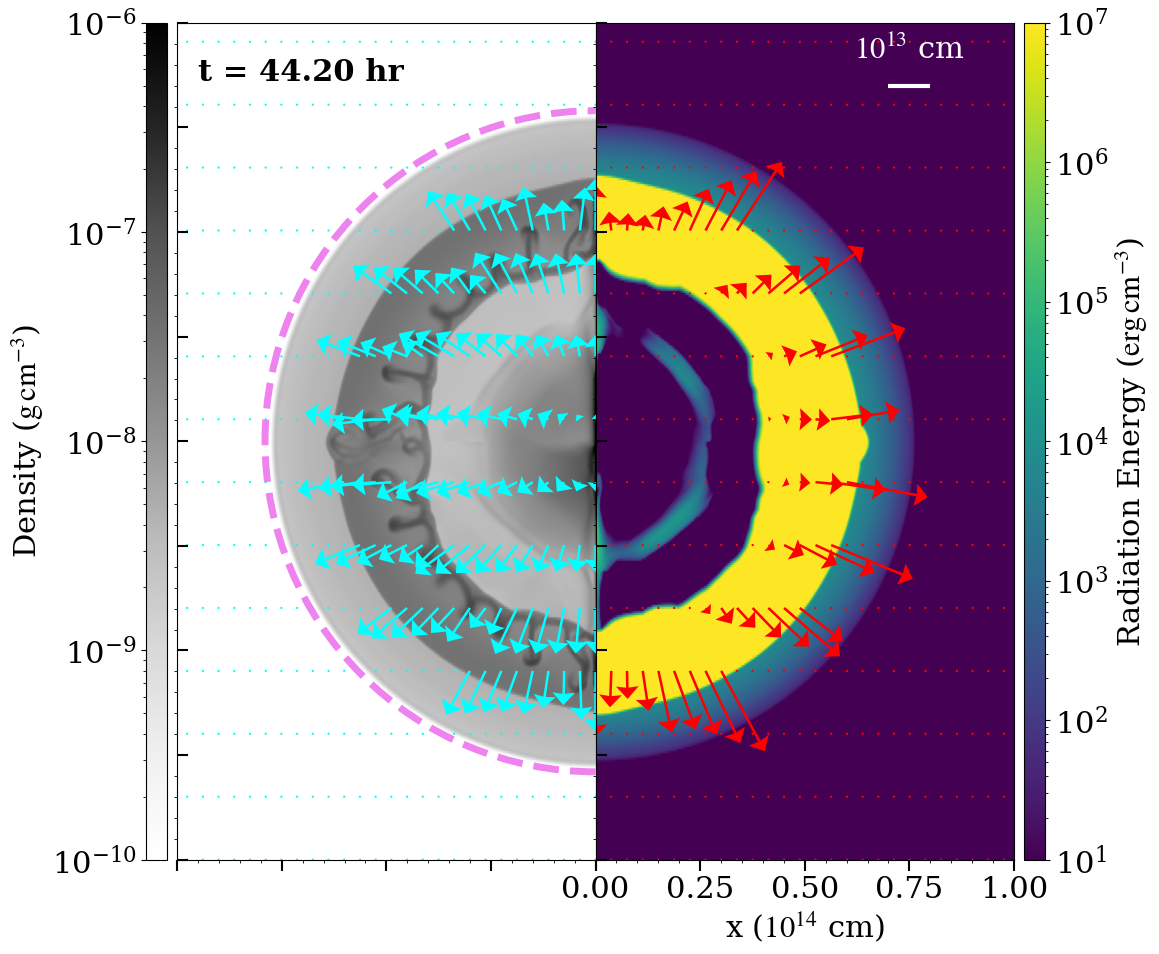}
    \caption{The snapshot of gas and radiation energy densities before the shock reaches the stellar surface for \mn{20}. Cyan and red vectors represent the velocity and radiation flux, respectively. The pink dashed line indicates the location of the photosphere. Rayleigh–Taylor fingers emerge around the contact discontinuity near ${\sim}4\times10^{13}$~cm. The atmosphere expands with a velocity that is negligible compared to the shock velocity. }
    \label{fig:verify}
\end{figure}

To ensure numerical stability, we introduce a buffer region with a power-law density profile extending from the stellar surface to the inner CSM, with a width of ${\approx}4 \times 10^{12}$~cm. This buffer places the stellar atmosphere slightly out of hydrostatic equilibrium, leading to a mild expansion. The left panel of \Figure{verify} shows that the envelope expands slowly with $v \le 10^4$~\cms, negligible compared to the shock velocity $v_{\rm s} \ge 10^8$~\cms. The luminosity of this expansion, shown in the right panel of \Figure{verify}, is negligible compared to the breakout radiation. Therefore, the buffer has only a minimal impact on modeling shock breakout.

We also find that the boundaries of the 2D simulation domain can introduce minor numerical artifacts, producing slightly enhanced velocities close to the $z$ axis. Nevertheless, these effects remain small and the global shock morphology remains nearly spherical.

\section{Results}
\label{sec:results}
\subsection{Gas Dynamics of Shock Breakout}
\label{subsec:gas}
\begin{figure}
    \centering
    \includegraphics[width=1.\textwidth]{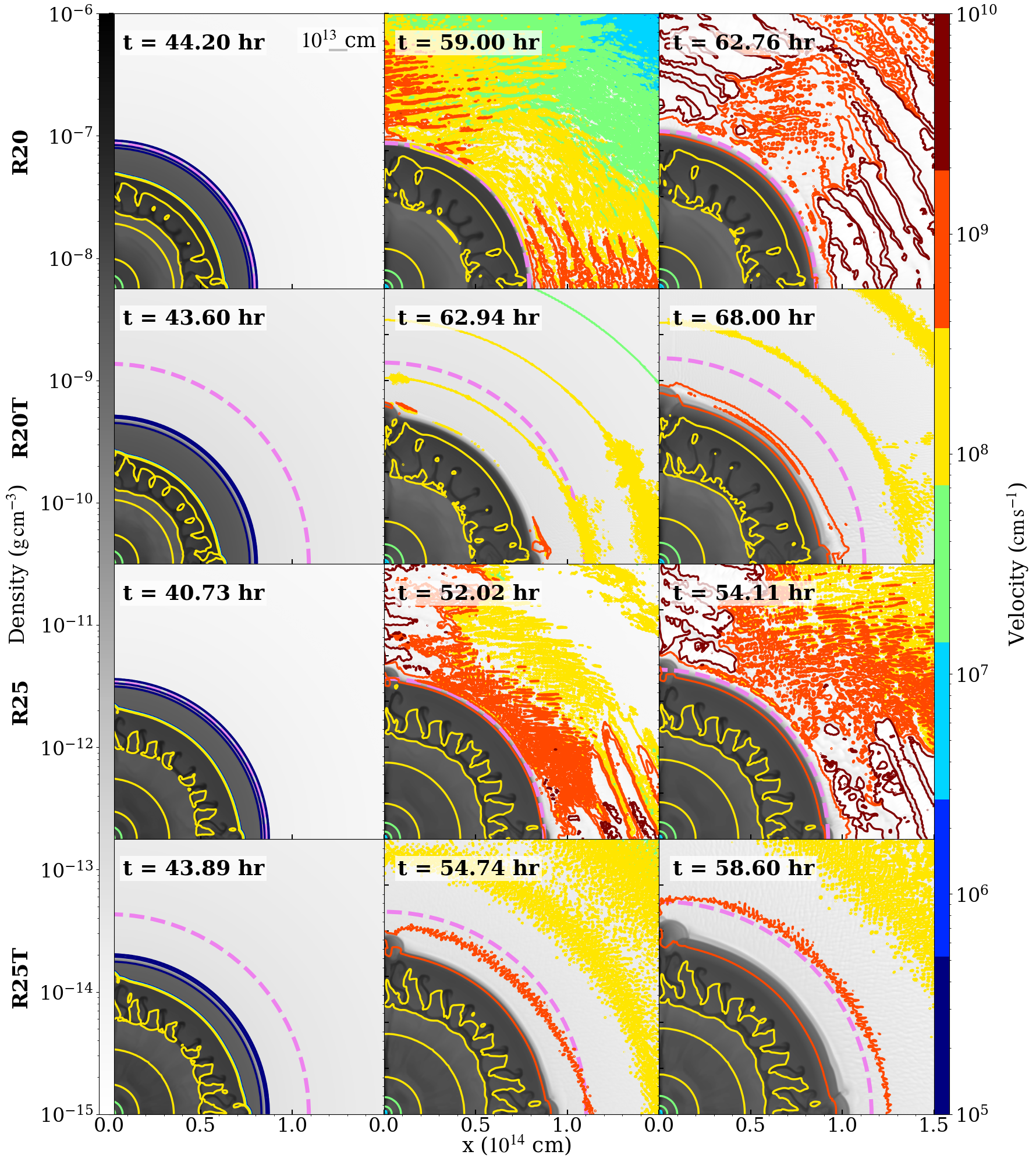}
    \caption{
    The evolution of the shock breakout for \mn{20}, \mn{20T}, \mn{25}, and \mn{25T} shown in each row. Each column, from left to right, represents breakout phases at pre-breakout, maximum luminosity, and post-breakout. The pink dashed line marks the photosphere. 
    In general, the stellar surface remains intact, but RT instabilities have developed within the envelope at the pre-breakout phase. When the breakout occurs and its emission reaches its peak, the velocity contour shows large velocity fluctuations in the CSM, indicating the RPS has started to develop and affects the locations of the photosphere. Some of the CSM can be accelerated by the radiation up to $v{\sim} 10^{10}$~\cms\ in \mn{20} and \mn{25} but $v{\sim}10^9$~\cms\ in \mn{20T} and \mn{25T}. The dense CSM can enlarge the radius of the photosphere and delay the timing of shock breakout. \rev{The velocity contours look noisy because of the small-scale velocities driven by the RHD instabilities.} }
    \label{fig:3}
\end{figure}

We first show the evolution of density and velocity from our models in \Figure{3}. As shown in the left column of \Figure{3}, the shock initially resides in optically thick regions well below the photosphere. The SN shocks start to break out and emit photons when they approach the stellar surface, where the local optical depth drops below $\tau\lesssim c/v_{\rm s}$. Right before the breakout occurs, the escaping radiation forms a radiation precursor (RP) ahead of the shock front, and it accelerates the overlaying CSM to ${\sim}$5--30\,\% speed of light depending on the explosion energy and CSM properties. Due to strong radiative cooling, the gas structure behind the RP is subject to fluid instabilities characterized by large density and velocity inhomogeneities. The resulting mixing alters the photosphere of shock breakout \citep{2008Schawinski,2024Chen}.

\begin{figure}
    \centering
    \includegraphics[width=1\linewidth]{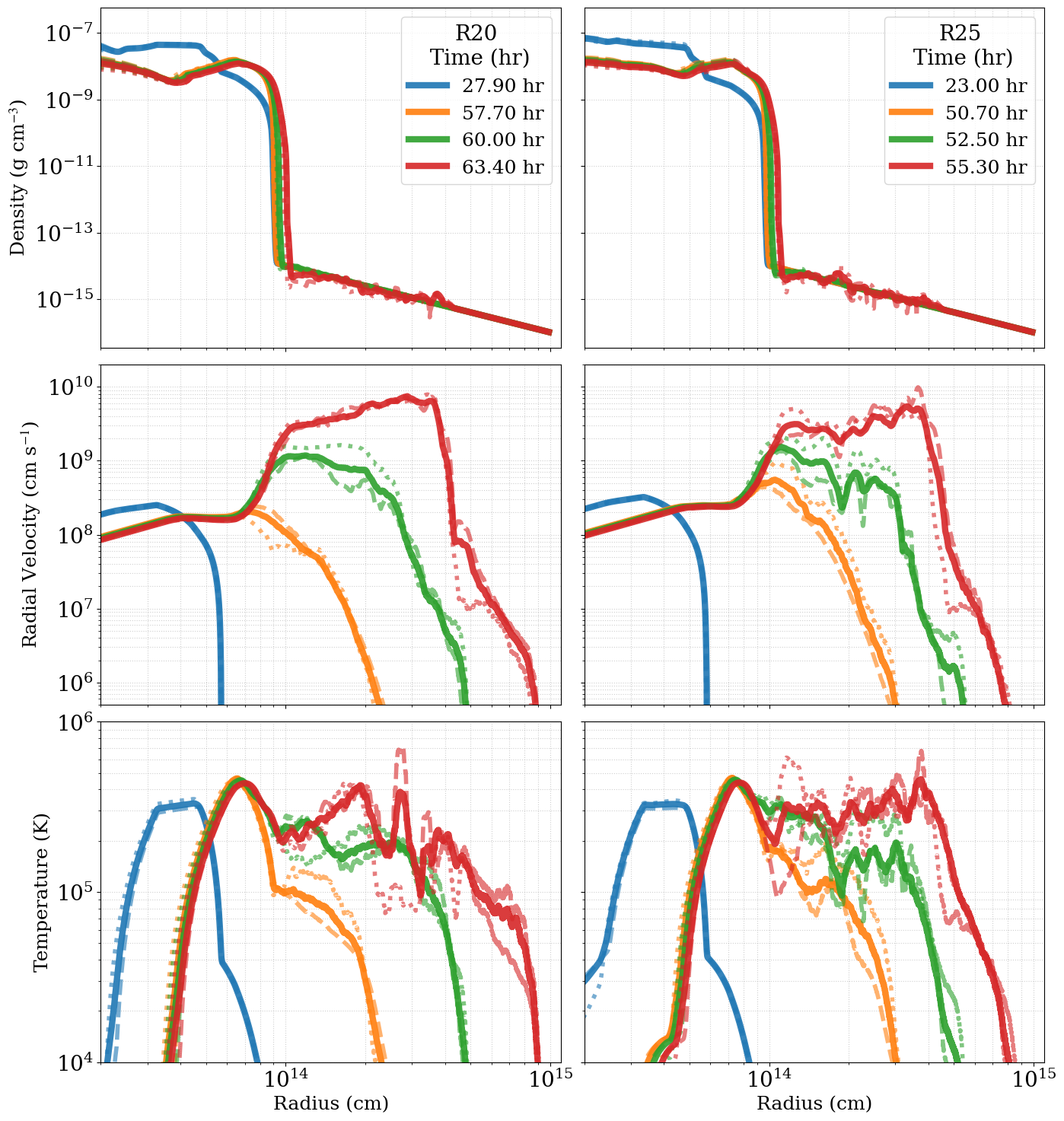}
    \caption{
    Evolution of the 1D density, velocity, and temperature profiles for \mn{20} (left) and \mn{25} (right) based on viewing angles of $15^\circ$ (dotted), $60^\circ$ (dashed), and angle-averaged (solid). Colors indicate key epochs: ${\sim}$30~hr pre-maximum luminosity (blue), ${\sim}$2~hr pre-maximum (orange), maximum luminosity (green), and post-breakout (red).
    As time evolves, angle-dependent profiles start to deviate significantly due to the developing RPS that causes large fluctuations in velocity and temperature of the CSM.}
    \label{fig:4}
\end{figure}

\begin{figure}
    \centering
    \includegraphics[width=1\linewidth]{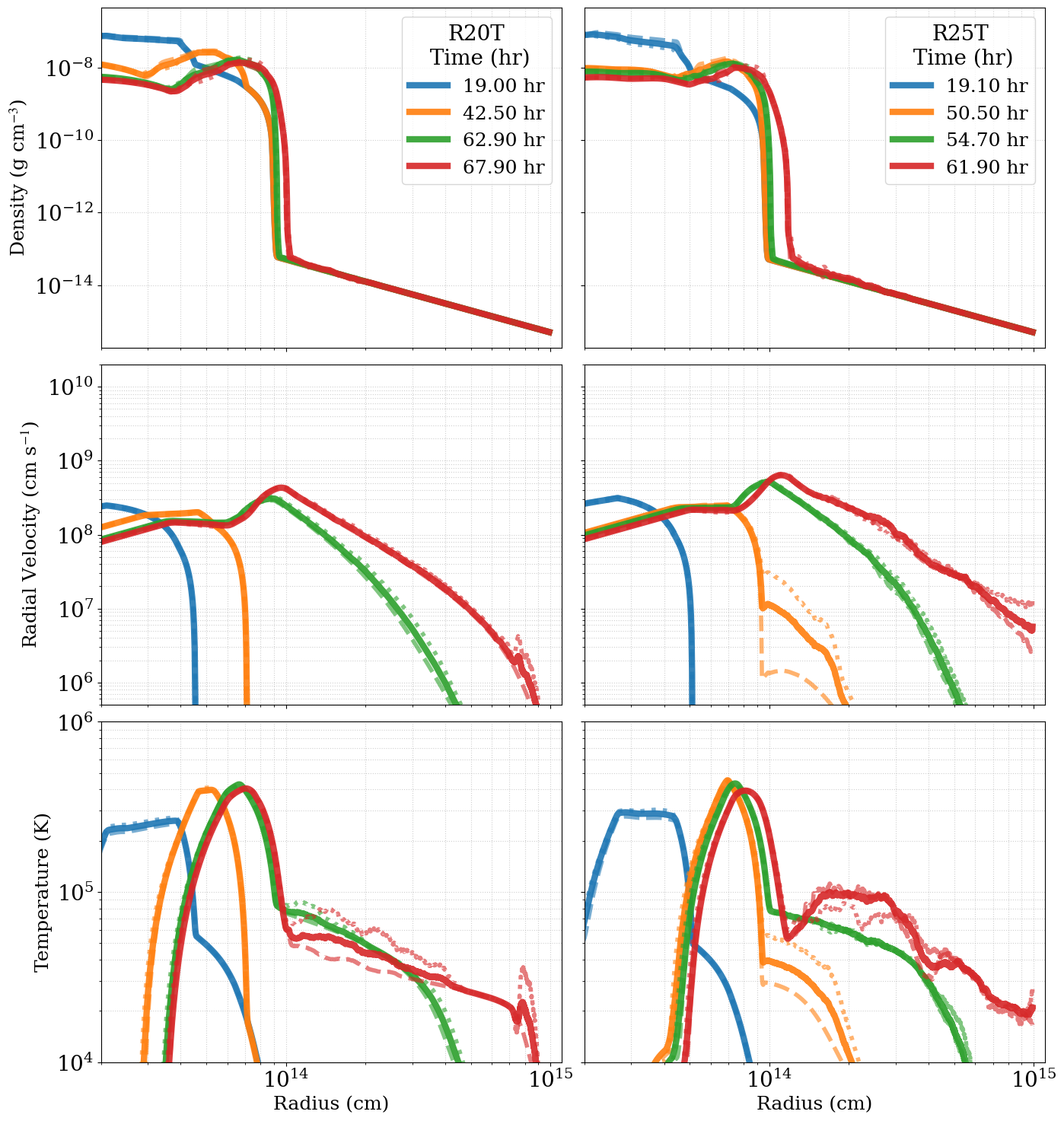}
    \caption{
Evolution of the 1D density, velocity, and temperature profiles for \mn{20T} (left) and \mn{25T} (right) based on viewing angles of $15^\circ$ (dotted), $60^\circ$ (dashed), and angle-averaged (solid). Colors indicate key epochs: ${\sim}$30~hr pre-maximum luminosity (blue), ${\sim}$2~hr pre-maximum (orange), maximum luminosity (green), and post-breakout (red). Similar to \Figure{4}, as time evolves, angle-dependent profiles start to deviate more due to the developing RPS that causes large fluctuations in velocity and temperatures of CSM. However, the amount of deviations in angle-dependent profiles of \mn{20T}/ \mn{25T} is less than that of \mn{20}/ \mn{25} due to the dense CSM. } 
    \label{fig:5}
\end{figure}

To examine the gas dynamics, we show the radial profiles of density, velocity, and temperature extracted from different viewing angles and epochs in \Figure{4} and \Figure{5}. Our 2D simulations exhibit smooth density transitions and temperature gradients across the forward shock (FS) compared to the initial shock structure shown in \Figure{1}.

Between the FS and the contact discontinuity, the emerging Rayleigh–Taylor fingers generate mild angular fluctuations. The RP accelerated structure (RPS) amplifies the inhomogeneous and anisotropic emission as the shock evolves. Approximately $2$~hr before reaching the maximum luminosity, the gas in RPS is accelerated to ${\gtrsim} 10^9$~\cms\ in models \mn{20} and \mn{25}, while in the dense CSM models \mn{20T} and \mn{25T}, the RPS velocities reach ${\sim}3\times 10^8$~\cms. The mixing produces clumpy structures in both the stellar atmosphere and the overlaying CSM, resulting in inhomogeneous optical depths and altering the photosphere properties. After the luminosity reaches its peak, the RPS inhomogeneities continue to evolve, and the FS begins to collide with the inner CSM. In summary, multidimensional RHD facilitates radiation leakage from FS, enhances RPS formation, and substantially alters shock breakout LCs similar to \citet{2024Chen}.

\subsection{LCs of Shock Breakout}
\label{subsec:LCs}
Shock breakout signals encode key information about the stellar radius ($R_{\rm *}$) and envelope properties of the SN progenitor star. A commonly estimated duration of shock breakout is the light-travel time $t\sim R_{\rm *}/c$ \citep{Katz_2010, 2017Lovergrove, Suwa2017lensing}. However, in reality, the shock breakout is affected by the envelope properties that make its emission a continuous process, rather than an instantaneous flash from the surface. It begins with photon leakage from behind the shock front at optical depths $\tau \sim c / v_{\rm s} \sim 10^2$--$10^3$, proceeds through the formation of the RPS at $\tau \sim 2/3$, and ends once the trapped radiation energy is released. The breakout duration is therefore governed by the optical depth between the explosion shock and the RPS. It is sensitive to hydrodynamic mixing, the extent of the stellar atmosphere, and the CSM density structure.

\begin{figure}
    \centering
    \includegraphics[width=1.0\linewidth]{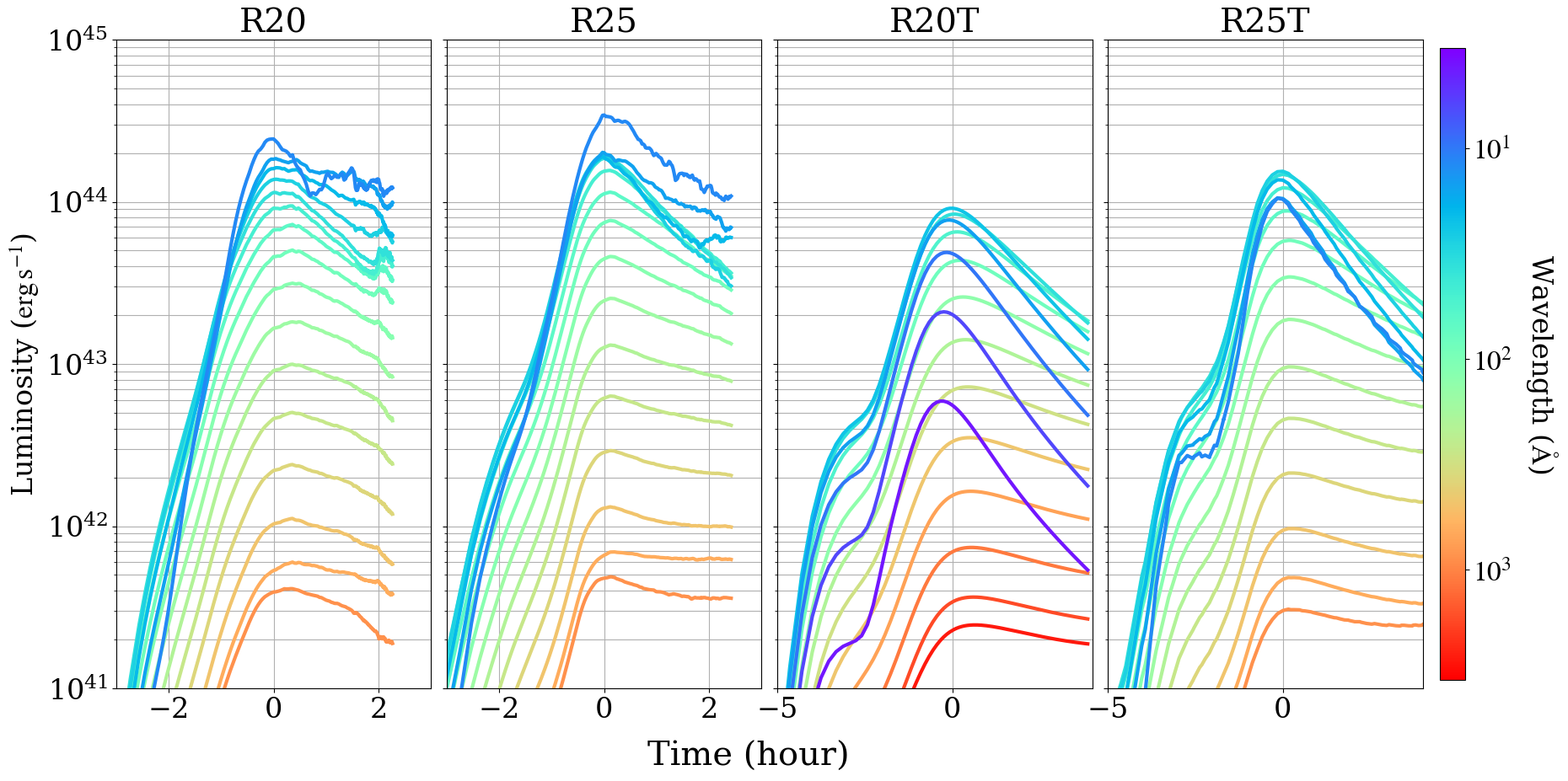}
    \caption{
    Multigroup LCs of all models at $\theta=45^{\circ}$. For comparison, the location of maximum luminosity has been shifted to $t=0$. Colors indicate the central wavelengths for each band. The luminosity starts to rise ${\sim}$2--4~hr before reaching its peak, where the bands with shorter wavelength radiation emerge and peak earlier than those of longer wavelengths. \mn{25} have higher luminosity compared to \mn{20}. The LCs of \mn{20T} and \mn{25T} are broader and less luminous than those of \mn{20} and \mn{25}. Due to a dense CSM, the post-maximum emission of \mn{20T} can be even overtaken by the red colors.}
    \label{fig:6}
\end{figure}

We present the color LCs from our simulations in \Figure{6}. To facilitate comparison among different models, the color LCs for a given model are temporally aligned with the time of the maximum luminosity. All light curves exhibit three distinct evolutionary phases: rise, peak, and decline. The breakout emission is dominated by ultraviolet radiation with characteristic wavelengths of ${\approx}50$--$200~\text{\AA}$. The maximum UV luminosity reaches ${\sim} (1$–$3)\times10^{44}\,\ergs$ for all models, with FWHM durations of approximately $1$--$3$ hr. The pre-shock blackbody emission (${\ll}10^{40}\,\ergs$) is negligible compared to the breakout luminosity (${\gtrsim}10^{43}$--$10^{44}\,\ergs$). The maximum luminosities and LC durations are summarized in Table~\ref{tab:2}.

In \Figure{6}, \mn{20} and \mn{25} both show a rapid rise in LCs and a gradual decay with FWHM durations of ${\sim}$1--3~hr. Short-wavelength (soft X-ray/UV) emission rises and decays more sharply, whereas radiation of longer wavelengths rises and decays more gradually. The maximum luminosities of \mn{25} exceed those of \mn{20} by ${\sim}$10--30\,\%, consistent with a larger $R_{\rm *}$ and explosion energy in \mn{25}. The durations of the FWHM of ${\sim}$1--3~hr are longer than the analytic estimates of around ${\sim}$0.5--1~hr \citep{Katz_2010, 2010Nakar}, but align well with multidimensional RSG breakout studies that include radiative diffusion effects \citep{2022Goldberg}.

For models with dense CSM, \mn{20T} and \mn{25T}, exhibit ${\sim}$50\,\% lower maximum luminosities, and they peak 2--4~hr later than \mn{20} and \mn{25}. LCs of \mn{20T} and \mn{25T} display smoother profiles and a mild pre-maximum shoulder at ${\sim}$2--3~hr before the major peak, driven by the heating of RPS and enhanced photon trapping in the dense CSM. The dense CSM prolongs effective diffusion time and shifts the emission toward longer wavelengths.

\begin{figure}
    \centering
    \includegraphics[width=\linewidth]{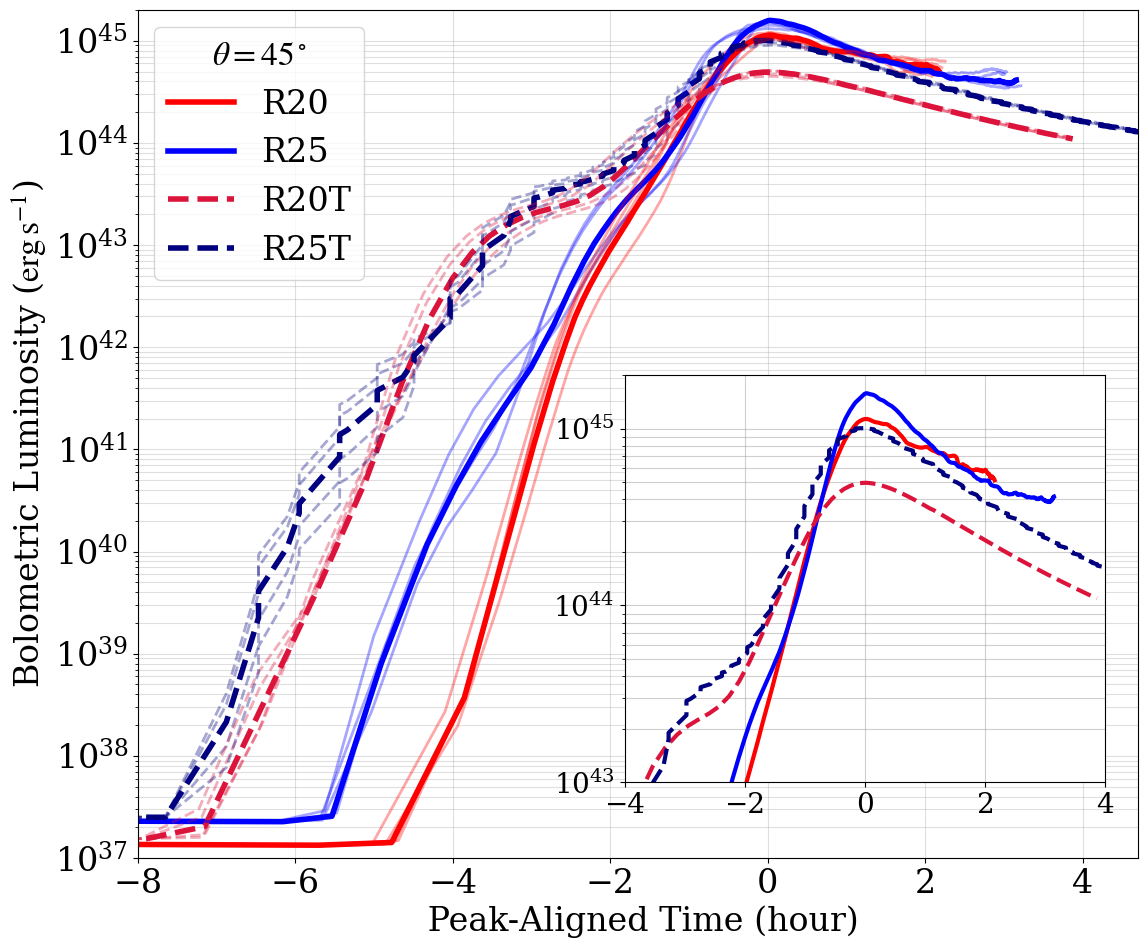}
    \caption{
    Bolometric LCs for all models viewed at $\theta = 45^\circ$, the semi-transparent lines within the same color show other viewing angles for the same model. For comparison, the location of maximum luminosity has been shifted to $t=0$. Models with a denser CSM exhibit earlier radiation heating, beginning ${\sim}$8~hr before reaching their maximum luminosity. The rise times to maximum luminosity are 4.8, 5.6, 7.1, and 7.7~hr for \mn{20}, \mn{25}, \mn{20T}, and \mn{25T}, respectively. The subplot shows the zoom-in around the maximum luminosities, showing the order of peak brightness as \mn{25} $>$ \mn{20} $>$ \mn{25T} $>$ \mn{20T}. }
    \label{fig:7}
\end{figure}

\Figure{7} presents the bolometric LCs of all models. The LCs computed from different viewing angles show minor variations, reflecting small-scale inhomogeneities caused by mixing from RPS. Interestingly, models with dense CSM exhibit extended LC profiles, characterized by an early-rising bump preceding the main peak. Overall, the multidimensional radiation–hydrodynamic simulations naturally reproduce the diversity of breakout LCs arising from variations in progenitor size and CSM density.

\begin{deluxetable*}{clcccccccc}
\tabletypesize{\scriptsize}
\tablecaption{
Characteristics of Multigroup LCs 
\label{tab:2}
}
\tablewidth{0pt}
\tablehead{
\multicolumn{2}{c}{} &
\multicolumn{2}{c}{R20} &
\multicolumn{2}{c}{R25} &
\multicolumn{2}{c}{R20T} &
\multicolumn{2}{c}{R25T} \\
{\begin{tabular}[c]{@{}c@{}}No. \\ ~\end{tabular}} & 
\colhead{\begin{tabular}[c]{@{}c@{}}Center$^{\rm *}$ \\ (\AA)\end{tabular}} & 
\colhead{\begin{tabular}[c]{@{}c@{}}Peak \\ (\ergs)\end{tabular}} & 
\colhead{\begin{tabular}[c]{@{}c@{}}FWHM \\ (hr)\end{tabular}} &
\colhead{\begin{tabular}[c]{@{}c@{}}Peak \\ (\ergs)\end{tabular}} & 
\colhead{\begin{tabular}[c]{@{}c@{}}FWHM \\ (hr)\end{tabular}} &
\colhead{\begin{tabular}[c]{@{}c@{}}Peak \\ (\ergs)\end{tabular}} & 
\colhead{\begin{tabular}[c]{@{}c@{}}FWHM \\ (hr)\end{tabular}} &
\colhead{\begin{tabular}[c]{@{}c@{}}Peak \\ (\ergs)\end{tabular}} & 
\colhead{\begin{tabular}[c]{@{}c@{}}FWHM \\ (hr)\end{tabular}}
}
\startdata
1 & 615.63 (1912.32)   & $9.96\times10^{42}$ & 2.72 & $1.31\times10^{43}$ & 3.51 & $7.22\times10^{42}$ & 5.63 & $9.60\times10^{42}$ & 7.29 \\
2 & 461.66 (1159.42)   & $1.82\times10^{43}$ & 2.70 & $2.54\times10^{43}$ & 3.06 & $1.41\times10^{43}$ & 5.08 & $1.88\times10^{43}$ & 5.03 \\
3 & 346.19 (702.94)  & $3.15\times10^{43}$ & 2.72 & $4.60\times10^{43}$ & 2.68 & $2.59\times10^{43}$ & 4.48 & $3.44\times10^{43}$ & 4.22 \\
4 & 259.61 (426.19)  & $5.04\times10^{43}$ & 2.77 & $7.67\times10^{43}$ & 2.25 & $4.36\times10^{43}$ & 3.86 & $5.78\times10^{43}$ & 3.71 \\
5 &194.68 (258.39)  & $7.25\times10^{43}$ & 2.29 & $1.15\times10^{44}$ & 2.04 & $6.52\times10^{43}$ & 3.39 & $8.79\times10^{43}$ & 3.26 \\
6 &145.99 (156.66)  & $9.44\times10^{43}$ & 2.09 & $1.56\times10^{44}$ & 1.89 & $8.39\times10^{43}$ & 3.10 & $1.22\times10^{44}$ & 2.92 \\
7 &109.48 (94.98)   & $1.14\times10^{44}$ & 2.10 & $1.86\times10^{44}$ & 1.65 & $9.12\times10^{43}$ & 2.74 & $1.48\times10^{44}$ & 2.50 \\
8 &82.10 (57.59)  & $1.38\times10^{44}$ & 2.17 & $1.92\times10^{44}$ & 1.48 & $7.72\times10^{43}$ & 2.67 & $1.54\times10^{44}$ & 2.34 \\
9 &61.56 (34.91)  & $1.62\times10^{44}$ & 2.60 & $1.90\times10^{44}$ & 1.54 & $4.87\times10^{43}$ & 2.46 & $1.37\times10^{44}$ & 2.07 \\
10 &46.17 (21.17)  & $1.85\times10^{44}$ & 2.69 & $2.02\times10^{44}$ & 1.89 & $2.10\times10^{43}$ & 2.32 & $1.06\times10^{44}$ & 2.01 \\
11 &34.62 (12.83)  & $2.44\times10^{44}$ & 1.23 & $3.43\times10^{44}$ & 1.59 & $5.90\times10^{42}$ & 2.37 & $1.05\times10^{44}$ & 1.84 \\
\enddata
\tablecomments{
Center$^{*}$ gives the central wavelength of each selected radiation group used for \mn{20}, \mn{25}, and \mn{25T}, while \mn{20T} adopts different values for the numerical stability shown in the parentheses.}
\end{deluxetable*}
\section{Discussion}
\label{sec:dis}
\subsection{Effect of Progenitor Stars}
\label{subsec:2}
\begin{figure}
    \centering
    \includegraphics[width=1\linewidth]{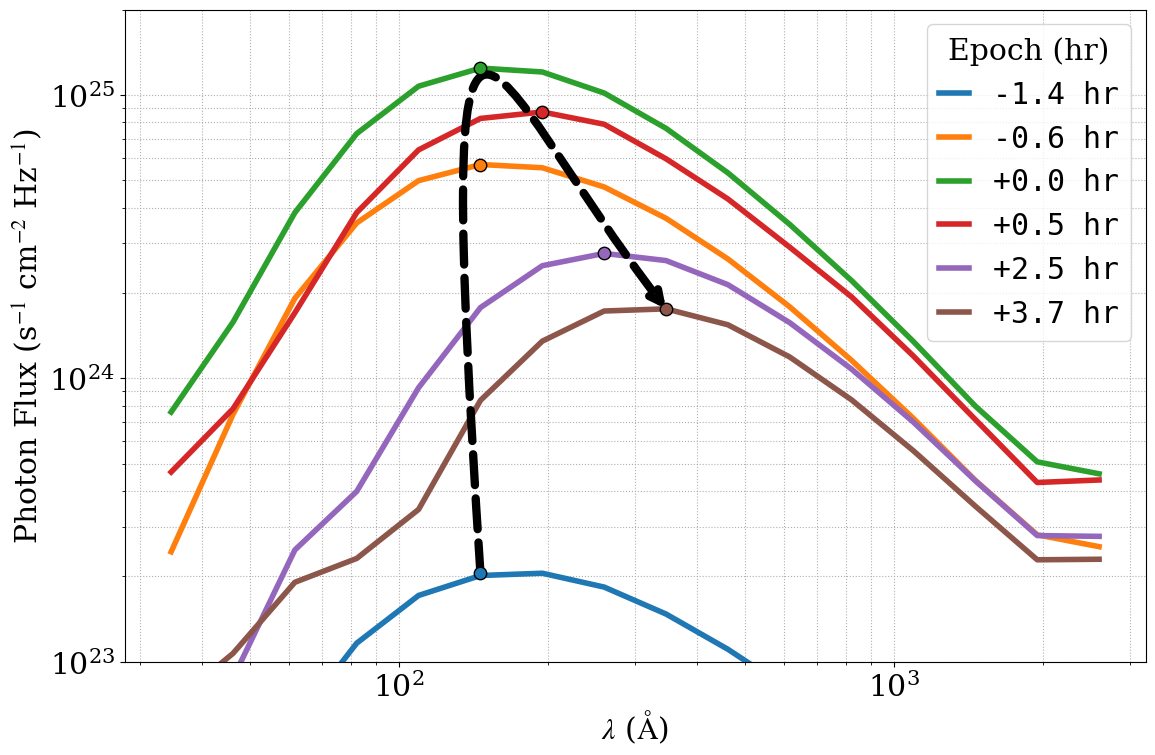}
    \caption{Evolution of spectral energy distribution (SED) for \mn{20} during the shock breakout. The SED is based on the radiation flux of different photon energy groups extracted at $r=9 \times 10^{13}$~cm at the viewing angle $\theta = 45^{\circ}$. Colored profiles represent different epochs of SED relative to the luminosity peak. We use the color circles to mark the location of peak flux for each SED, and a black dashed line with an arrow shows the evolution of SED peaks at different times. Before the luminosity peak, the peak flux of $\lambda\sim120$--$200\,\text{\AA}$ rises sharply in flux within ${\sim}1$ hr. After the luminosity peak, the flux peak of SED shifts to longer wavelengths of $\lambda \sim200$--350~$\text{\AA}$ and the magnitude of peak flux decreases gradually. }
    \label{fig:8}
\end{figure}

As shown in \Figure{6}, the LCs of shorter-wavelength ($\lambda \leq 100\,\text{\AA}$) evolve more rapidly than those of longer-wavelength ($\lambda \gtrsim 300\,\text{\AA}$) because the high-energy photons from the FS quickly heat the CSM to form RPS and shift redder. Furthermore, the longer-wavelength radiation later streams from the cooling of the shocked envelope, which prolongs LCs. We also show the evolution of the spectral energy distribution (SED) of \mn{20} in \Figure{8}. Before approaching the peak luminosity, the wavelength of dominant photon energy stays at $\lambda \sim 100$--$150\,\text{\AA}$ and the corresponding flux increases by a factor of 100 to the peak within 1.5~hr. After passing the peak luminosity, the wavelength of dominant photon energy gradually shifts to ${\sim} 300\,\text{\AA}$ and the corresponding flux decreases by a factor of 10 compared to the peak within 3.7 hr.

Comparing the color LCs of shock breakout helps clarify the explosion energy and the stellar structure of progenitor stars \citep{2022Goldberg}. As shown in \Figure{7}, the bolometric LCs of \mn{25} are approximately 30\,\% brighter than those of \mn{20}, reflecting its higher explosion energy. Furthermore, Table~\ref{tab:2} indicates that the shorter-wavelength emission of \mn{20} ($30 \lesssim \lambda \lesssim 100~\text{\AA}$) persists longer, whereas its longer-wavelength emission ($\lambda \gtrsim 300~\text{\AA}$) declines more quickly compared to \mn{25}. These differences arise from both the explosion energy and the structure of the hydrogen envelope.

Studies by \citet{2010Yoon, 2017Fuller} imply that the eruptive mass loss of massive stars can significantly alter the RSG structure prior to the onset of SNe. For example, steep density gradients in the atmosphere produce sharper, more luminous breakouts, while shallower gradients produce a dimmer emission \citep{2017Lovergrove, Ou_2023}. If an RSG contains an extensive and optically thin layer attached to its original envelope, it can also prolong the breakout LCs by increasing the photon diffusion time \citep{2015Moriya}. Furthermore, the clumpy envelope can significantly alter the signatures of shock breakouts. Recent 3D RSG models show that large-scale convective motion in the envelope and the generation of clumpy structures can increase the durations of the breakout \citep{2022Goldberg}. Our time-resolved multigroup LCs offer diagnostics to break down these structural degeneracies into the observables of shock breakout.

Furthermore, our RSG models exhibit distinct shock breakout signatures compared to those of Wolf–Rayet (WR) stars and blue supergiant (BSG) models with CSM interaction. WR breakouts, such as SN 2008D, show durations of ${\sim}$0.1~hr in the presence of dense CSM \citep{2010Couch, 2011Couch,2013Sapir,2014Svirski}, while BSG models produce ${\sim}$0.5~hr X-ray peaks \citep{2024Chen}. For example, the SN 1987A progenitor, with a radius of $3 \times 10^{12}$ cm and a shock width of ${\sim} 10^{12}$ cm, yields maximum luminosities ${\sim} 10^{46}$~\ergs\ and FWHM durations of ${\sim}$0.5~hr \citep{1992Ensman, 2024Chen}. In contrast, breakouts in RSGs are typically longer and dimmer due to their extended, low-density envelopes, which cause the shock to decelerate and heat up the envelope over a longer timescale \citep{Katz_2010}. Observational constraints from \citet{2008Schawinski} show UV breakout luminosities of ${\sim}10^{44}$~\ergs\ with durations of several hours, in contrast to compact progenitors such as SN 2008D that exhibit X-ray bursts of ${\sim}10^{46}$~\ergs\ over ${\sim}$100 seconds \citep{2008Soderberg}. Due to an extensive hydrogen envelope in RSG, the breakout emission is converted from X-ray to UV, and the LC duration extends to several hours. 

\subsection{Effect of CSM Profiles}
\label{subsec:3}
RSGs typically undergo substantial mass loss before the iron core collapse. In some cases, multiple giant eruptions expel part of the envelope shortly before the explosion, producing dense CSM that can alter the observed SN LCs \citep{2015Gezari, 2016Graefener}. Previous 1D RHD studies have shown that shock and dense CSM interactions can prolong the LCs of shock breakout and produce higher luminosities due to a large photosphere radius with more radiation energy through shock heating \citep{2010Ofek,2011Chevalier}. To fit with the late-time shock breakout observations, 1D models usually require an extremely-high mass-loss rate of $\dot{M}\sim10^{-2}\,\Ms\,\rm{yr^{-1}}$ \citep{2022Margalit,2025Hu}. \rev{1D models of \citet{2017Dessart} demonstrate that such winds can produce redder, broader LCs with pre-maximum shoulders. In contrast, our multidimensional simulations show that RPS and mixing naturally enhance the effective photon diffusion time, yielding similarly extended breakout durations and morphological features without requiring such extreme wind densities. Consequently, 1D models that neglect multidimensional mixing tend to underestimate LC durations and overestimate peak luminosities \citep[see discussion in][]{2022Goldberg}.}

\rev{A dense CSM increases the photospheric radius; meanwhile, it absorbs some fraction of the radiation energy from the RP to drive its expansion. As a result, the total escaping radiation is reduced due to the higher effective optical depth, leading to a lower peak luminosity during shock breakout.}  

In our multidimensional simulations, we assume a spherically symmetric dense CSM structure for models \mn{20T} and \mn{25T}. As shown in \Figure{7}, comparing \mn{20} and \mn{25} with their CSM counterparts (\mn{20T} and \mn{25T}) reveals that models with dense CSM produce broader and dimmer LCs. For example, the photospheric radii of \mn{20} and \mn{20T} are $8.1\times10^{13}$~cm and $1.1\times10^{14}$~cm, respectively. The dense CSM also reduces the photospheric temperature from $T \sim 2\times10^{5}$~K in \mn{20} to ${\sim} 6\times10^{4}$~K in \mn{20T}. Consequently, the LCs of \mn{20T} are dominated by longer-wavelength emission relative to those of \mn{20}, as shown in \Figure{6}.

The LC durations can be estimated from the shock crossing time at the breakout radius \citep[e.g.,][]{2017Waxman}:
\begin{equation}
    t \sim \frac{R}{v} \sim \frac{l\,\tau}{v} \sim \frac{c}{\kappa \rho v^2}\rm{,}
\end{equation}

where $v$ is the gas velocity in the photosphere, $\tau = c/v$ is the optical depth, and $l = 1/(\kappa\rho)$ is the mean free path of the photon. This yields a breakout duration of ${\sim} 0.61$~hr for \mn{20} and ${\sim}$0.72~hr for \mn{20T}, corresponding to a ${\sim}$20\,\% increase. This trend is consistent with the averaged durations of color LCs, which are $2.37$~hr for \mn{20} and $3.12$~hr for \mn{20T}.

A similar comparison between \mn{20T} and \mn{25T} gives estimated breakout durations of $0.72$~hr and $0.8$~hr, respectively. Again, this agrees with the averaged duration of color LCs, which increases from $3.12$~hr in \mn{20T} to $3.34$~hr in \mn{25T}. 

The RPS modifies the breakout timescale by early RP leakage at the rising phase and increases the effective optical depth through CSM mixing. To quantify this effect on the bolometric LCs (\Figure{7}), we perform a linear fit relating the breakout rise time, estimated from the onset and the peak of breakout emission, to the RSG stellar radius and its mass-loss rate
\begin{equation}
t \approx 4.63 + 0.75 \left( \frac{\dot{M}}{6\times10^{-6}\,\Ms\,\mathrm{yr}^{-1}} \right) \cdot \frac{R_{\rm *}}{c} \quad {\rm hr}\rm{,}
\end{equation}
where $R_{\rm *}/c$ is the LTT in hours. The intercept (4.63~hr) accounts for the dynamical and geometric contributions, while the coefficient 0.75 represents the increase in diffusion time due to the denser CSM.
 
Our 2D LCs agree with the shock breakout observation in terms of LC durations, for example, SNLS-04D2dc, PS1-13arp, and SN 2016gkg \citep{2008Schawinski, 2015Gezari, 2018Bersten}. In contrast, previous 1D models would require orders-of-magnitude higher mass-loss rates for creating a dense CSM \citep{Ginzburg2014, 2018Foerster} or tailored stellar structures with an inflated envelope or convective atmosphere \citep{2015Moriya, 2022Goldberg} to yield similar durations.

\subsection{Important Caveats}
Multigroup simulations offer observational diagnostics for identifying extended breakout signatures of RSG explosions. Although our 2D simulations mitigate the mixing deficiency present in 1D models, which often produces an artificially thin shell at the shock front, their dynamics still differ from true 3D behavior, where turbulent energy cascades from large to small scales \citep{Ken2023ApJ}. However, 3D effects have a limited impact on shock breakout signals without highly asymmetric explosions. Our 2D results are consistent with 3D shock breakout simulations considering the clumpy envelope \citep{2022Goldberg}. \rev{We note that the FLD approximation adopted in this study may become less accurate near the photosphere and in optically thin regions, which represents one of the main caveats of our approach.}

We adopted simple explosion conditions by using 1D spherical progenitor stars and explosion models, as well as constant stellar winds for CSM. In reality, the explosion can be highly asymmetric with clumpy CSM as suggested by \cite{2017Yaron,2025Chen}. Moreover, the constant electron scattering opacity may be oversimplified when radiation is transported to cooler, denser CSM regions, where hydrogen recombination and gas metallicity may significantly influence bound-bound and bound-free absorptions.

\rev{Although our multigroup light curves primarily cover the extreme-UV to soft X-ray regime, they capture the critical photospheric cooling phase that bridges current and future wide-field surveys. In particular, our simulations resolve the spectral peak shifting from $\sim 100$~\AA\ toward longer wavelengths (\Figure{8}). Current and upcoming space observatories such as Einstein Probe (EP; $\sim$3--25~\AA; \citealt{2025Yuan,2025Cheng}) and ULTRASAT ($\sim$2300--2900~\AA; \citealt{2024Shvartzvald}) will be able to probe shock breakouts in great detail. Unfortunately, our selected energy ranges fall outside the observational bands of EP and ULTRASAT, preventing direct quantitative predictions for these missions. Nevertheless, our study provides valuable insights into the color evolution of shock breakout and can still help guide the interpretation of future observations.}

\rev{Our multidimensional simulations demonstrate that fluid instabilities of the RPS maintain a prolonged effective diffusion time and alter color evolution. This implies that inferring progenitor properties by fitting 1D templates to observations may introduce systematic biases \citep{2025Srinivasaragavan}. Our results therefore highlight the necessity of using multidimensional RHD models to correctly interpret the multi-band time delays and spectral energy distributions of events like EP250827b \citep{2025Sun}.}

\section{Conclusions}
\label{sec:conclusions}
We present the first 2D multigroup RHD simulations of RSG shock breakouts for 20~\Ms\ and 25~\Ms\ solar-metallicity progenitors, explicitly accounting for radiation precursor structures (RPS) across a range of CSM densities. The multidimensional geometry greatly improves the treatment of RPS by enabling more realistic radiation transport and mixing processes that are absent in 1D calculations. Our results show that strong radiation precursors originating from radiation leakage behind the explosion shock can drive fluid instabilities and shift the photosphere location prior to shock emergence. 

The resulting shock breakout emissions reach peak luminosities of ${\sim} 10^{44}$~\ergs\ with full-width half-maximum durations of 1--3~hr, substantially dimmer and longer than those from blue supergiant progenitors, which typically peak at ${\sim} 10^{46}$~\ergs\ with durations of ${\sim} 0.5$~hr. The corresponding LC colors evolve gradually from blue to red after the peak. We find that the 25~\Ms\ model with an explosion energy of $E \sim 1.69\times10^{51}$~erg produces ${\sim}$10--30\% higher peak luminosities than the 20~\Ms\ model with $E \sim 1.09\times10^{51}$~erg. If a dense CSM surrounds the progenitor, the rising time of the breakout luminosity is extended due to the increased photon diffusion time in the denser medium.

Our results also demonstrate that multidimensional effects make shock breakout much more sensitive to CSM properties than predicted by previous 1D or analytic models, which typically yield rise times of ${\lesssim}$1~hr for RSGs \citep{2017Lovergrove, 2022Margalit, 2022Goldberg}. The LCs from our 2D simulations show better agreement with observations than previous 1D models, which often require invoking extreme values of $\dot{M} \gtrsim 10^{-3}\,M_\odot\,{\rm yr}^{-1}$ to account for delayed breakout signals \citep{2016Tanaka, 2018Foerster, 2023Hosseninzadeh}. Therefore, previous 1D models likely overestimate the required mass-loss rate to reproduce the observed breakout LCs.

Our results provide new insights into the radii and atmospheric structures of red supergiants, offering a promising avenue for probing their mass-loss histories and the pre-explosion conditions of massive stars. Future work will explore more complex CSM geometries and perform fully three-dimensional RHD simulations with improved opacity treatments to further refine our understanding of shock breakout physics and the diverse circumstellar environments associated with massive star explosions.

\begin{acknowledgments}
This research is supported by the National Science and Technology Council, Taiwan, under grant No. MOST 110-2112-M-001-068-MY3, NSTC 113-2112-M-001-028-, 114-2811-M-001-094, and the Academia Sinica, Taiwan, under a career development award under grant No. AS-CDA-111-M04. KC acknowledges the support of the Alexander von Humboldt Foundation. WC acknowledges support of the National Taiwan University Scholarship for Direct Pursuit of Doctoral Degree, the Scholarship of the Chung Hwa Rotary Education Foundation (Rotary International District 3490, Sanchung–Sanyang Rotary Club), and the 2025 National Science and Technology Council-Deutscher Akademischer Austauschdienst (NSTC–DAAD) Summer Institute Programme. KM acknowledges support from the Japan Society for the Promotion of Science (JSPS) KAKENHI grant Nos. 24KK0070 and 24H01810. 
Our computing resources were supported by the National Energy Research Scientific Computing Center (NERSC), a U.S. Department of Energy Office of Science User Facility operated under Contract No. DE-AC02-05CH11231, and the KAWAS Cluster at the Academia Sinica Institute of Astronomy and Astrophysics (ASIAA). The software, {\FLASH}, used in this work was developed in part by the DOE NNSA- and DOE office of Science-supported Flash Center for Computational Science at the University of Chicago and the University of Rochester. The work of FKR  is supported by the Klaus Tschira Foundation, by the Deutsche Forschungsgemeinschaft (DFG, German Research Foundation) -- RO 3676/7-1, project number 537700965, and by the European Union (ERC, ExCEED, project number 101096243). Views and opinions expressed are, however, those of the authors only and do not necessarily reflect those of the European Union or the European Research Council Executive Agency. Neither the European Union nor the granting authority can be held responsible for them. 
\end{acknowledgments}
\bibliography{sample631}
\bibliographystyle{aasjournalv7}
\end{CJK*}
\end{document}